\documentclass[fleqn,usenatbib]{mnras}

\usepackage[T1]{fontenc}
\usepackage{ae,aecompl}

\DeclareRobustCommand{\VAN}[3]{#2}
\let\VANthebibliography\thebibliography
\def\thebibliography{\DeclareRobustCommand{\VAN}[3]{##3}\VANthebibliography}

\usepackage{color,soul}
\usepackage{graphicx}	
\usepackage{amsmath}	
\usepackage{amssymb} 
\usepackage{moresize}
\usepackage{multicol}
\usepackage{booktabs,tabularx}
\usepackage{enumitem}
\usepackage{multirow}
\usepackage{adjustbox}
\usepackage{gensymb}
\usepackage[toc,page]{appendix}

\newcommand{\msun}{\text{M}_\odot} 
\newcommand{\lsun}{\text{L}_\odot} 

\title[On the Progenitor of the Crab Pulsar]{On the Progenitor of the Crab Pulsar}

\author[Cruz-Cruz \& Kochanek]{
Elvira Cruz-Cruz,$^{1}$\thanks{E-mail: cruz-cruz.1@osu.edu (ECC)} and
C.~S.~Kochanek,$^{1,2}$
\\
$^{1}$Department of Astronomy, The Ohio State University, 140 West 18th Avenue, Columbus, OH 43210, USA \\
$^{2}$Center for Cosmology and AstroParticle Physics (CCAPP), The Ohio State University, 191 W. Woodruff Avenue, Columbus, OH 43210, USA\\
}

\date{Accepted XXX. Received YYY; in original form ZZZ}

\pubyear{2022}

\begin{document}
\label{firstpage}
\pagerange{\pageref{firstpage}--\pageref{lastpage}}
\maketitle

\begin{abstract}
 The Crab supernova is interesting because we know that it was not a binary at death, the outcome was a neutron star, and because of the supernova remnant's apparently low energy and mass. Using Gaia EDR3 parallaxes and photometry, we examine the stellar population local to the Crab in a cylinder with a projected radius of 100 pc and parallax range $0.427 < \varpi < 0.619$ mas set by the uncertainties in the Crab's parallax. We also individually model the most luminous stars local to the Crab. The two most luminous stars are blue, roughly main sequence stars with masses of $\sim11 M_{\odot}$. We estimate the stellar population's age distribution using Solar metallicity \texttt{PARSEC} isochrones. The estimated age distribution of the 205 $M_{G} < 0$ stars modestly favor lower mass stars consistent with an AGB star or a lower mass binary merger as the progenitor, but we cannot rule out higher masses. This may be driven by contamination due to the $\sim700$ pc length of the cylinder in distance.

\end{abstract}




\section{Introduction}
We need to understand the progenitors of core-collapse supernovae (ccSNe) to understand the final stages of massive stellar evolution and their relation to supernovae and their remnants. Understanding the deaths of massive stars is crucial for comprehending their evolution, the role of binaries, and the origins of the systems that merge and produce gravitational waves. In particular we would like to know the masses of SN progenitors. Three methods have been used to constrain progenitor masses: (1) direct observations of progenitors, (2) X-ray estimates of supernovae remnant (SNR) compositions; and (3) analyses of the stellar populations local to the supernovae. The latter two methods are indirect but have the advantage that they can be used long after the explosion occurred.

The progenitors of Type II-P SNe are the best constrained thanks to direct observational detections of their progenitors in (mostly) archival HST data \citep[e.g.,][]{ Smarttetal2002a, Smarttetal2002b, Smarttetal2009a, VanDyk2003a, VanDyke2012b, Lietal2006, Lietal2007, Maundetal2005, Maundetal2011, Maundetal2014, Hendryetal2006, Smithetal2011, Fraseretal2014}. \citet{Smarttetal2009a},  using 8 mass estimates and 12 upper limits for Type IIP SN progenitors found a minimum mass of $M< 8.5^{+1.0}_{-1.5} M_{\odot}$ and a maximum mass of $M_{max} = 16.5 \pm1.5 M_{\odot}$ assuming a Salpeter IMF. The progenitors to Type IIP supernovae are all red supergiants (RSG) \citep{Smarttetal2009a}. In a later review of 18 mass estimates and 27 upper limits, \citet{Smartt2015} found an upper mass limit for RSGs exploding as Type II SN of about $18 M_{\odot}$. Since stellar models predict that RSGs of up to $30 M_{\odot}$ undergo core-collapse and could produce Type II SN, the missing $18-30 M_{\odot}$ progenitors has been termed the red supergiant problem. Since then there has been ongoing debate about the existence of this mass range problem \citep[e.g.,][]{Kochanek2012, KochanekRSG2020, Beasor2020, Walmswell&Eldridge2012, Groh2013, Davies&Beasor2018, Davies&Beasor2020, Strotjohann2024}. There are fewer direct detections of progenitors to Type Ibc supernovae because stripped stars tend to be optically faint and difficult to detect  \citep[e.g.,][]{Eldridgeetal2013, Folatellietal2016, Johnsonetal2017TypeIbc, Kilpatricketal2021TypeIb, Yoonetal2012}.

A second method for understanding supernovae and their progenitors is to analyze the X-ray emission from the ejecta. \cite{Katsudaetal2018} made progenitor mass estimates for $33$ core-collapse SNRs in our Galaxy and the Small and Large Magellanic Clouds, focusing on the Fe/Si abundance ratio. They argue that the Fe/Si ratio is the best estimate of the progenitor's CO core mass and thus the initial progenitor mass $M_{ZAMS}$. \cite{Katsudaetal2018} splits the sample into three mass bins where $M_{ZAMS} < 15 M_{\odot}$, $15M_{\odot} < M_{ZAMS} < 22.5M_{\odot}$, and $22.5M_{\odot} < M_{ZAMS}$ to model the bin fraction with and without a mass cutoff. They argue that the observed distribution better agrees with models lacking the mass cut off implied by the red supergiant problem.

The third method is to analyze the stellar populations near both SNe and SNRs in external galaxies. The color-magnitude diagram (CMD) of the nearby stars is modeled with isochrones to derive the local star formation history, which then provides a probability distribution for the mass of the star which exploded. For example, \cite{Jenningsprogmassdistr2014} found supernovae remnant progenitor mass distributions for M31 and M33, \cite{Auchettl2019} did so for the Small Magellanic Cloud, \cite{Murphyprogmass2011} did so for SN 2011dh, and \cite{Williamsprogmass2014} \citep{Diazprogmassdistr2021} compiled progenitor mass constraints for 17 (22) historic core-collapse supernovae. \cite{Diazprogmassdistr2021}, for example, finds a progenitor mass distribution with a minimum mass of $M_{min} = 8.60^{+0.37}_{-0.41} M_{\odot}$ and a slope of $\alpha = -2.61^{+1.05}_{-1.18}$.


Gaia \citep{GaiaCollab2016,GaiaCollab2021} makes applying the stellar population analysis method feasible in our Galaxy \citep{velapulsar2022}. Galactic SNRs have the advantage that we frequently know the result of the explosion and the binarity of the progenitor \citep[e.g.,][]{Ilovaisky1972GalacticSNRs, Boubert2017binarycompanions,NotBinariesatdeathKochanek2018, Fortin2024FLMXBsGalactic}. Accurate parallaxes allow both the selection of stars local to the SNR and the determination of their luminosities. Three dimensional dust maps \citep[e.g.,][]{Bovy2016,Green2019} enabled by Gaia help to constrain the individual stellar extinctions. \cite{velapulsar2022} and \cite{Murphy2024VelaPaperBinary} successfully applied this method to the Vela pulsar. \cite{velapulsar2022} found a progenitor mass estimate of $\leq 15 M_{\odot}$ and \cite{Murphy2024VelaPaperBinary} found evidence that Vela's progenitor was the product of a binary merger. Suitable candidates for this method do require a well-constrained SNR distance which is frequently a problem, although \cite{Kochanek2024S147} demonstrated a method which should provide accurate distances to any SNR with modest extinction. One good candidate is the Crab SNR where Gaia DR3 \citep{GaiaCollab2021} and VLBI observations of the pulsar \citep{LinRebeccaVLBI2023} provide well-measured distances, we know that the outcome of the explosion was a neutron star \citep[e.g.,][]{Staelin&Reifestein1968,Comella1969}, and that the system was not a binary at death \citep{NotBinariesatdeathKochanek2018}.

The origin of the Crab SNe (i.e, ccSNe or electron capture SNe), and remnant, has long been a topic of discussion \citep[e.g.,][]{Clark&Stephenson1977, Davidson&Fesen1985, Collinsetal1999PASP}. The Crab nebula and pulsar are the remnants of SN 1054 \citep[e.g.,][]{Duyvendak1942, Mayall&Oort1942, Staelin&Reifestein1968, Comella1969}. The SNR appears to be low mass and have a low kinetic energy of $\approx10^{49}$ erg, lower than the expected kinetic energy of a core-collapse supernova $\approx 10^{51}$ erg, \citep[e.g.,][]{MacAlpineetal1989, Bientenholz&Kronbergetal1991, Fesen&Shulletal1997, Smith2003}. The SN was a very luminous event with a peak absolute visual magnitude of $-18$ mag \citep[e.g.,][]{Chevalier1977, Trimble1973, Miller1973} that is brighter than typical Type II ccSNe \citep[e.g.,][]{Li2011peakluminoisity}. \cite{crabnebula2013} argues that the Crab was a Type IIn-P supernova caused by a sub-energetic electron-capture explosion of an $8-10 M_{\odot}$ super-AGB star. Electron capture SNe are generally associated with the mass range of extreme AGB stars \citep{Miyajietal1980, Nomotoetal1982,Nomotoetal1987}, although the exact mass range depends on the model (e.g., \citet[][]{Poelarends2008} and \citet{Limongi2024} find $9.00-9.25 M_{\odot}$ and $8.5-9.2 M_{\odot}$, respectively). They are predicted to be underluminous and underenergetic \citep[e.g.,][]{Kitaura2006}. In the \cite{crabnebula2013} scenario, the high luminosity is not driven by the normal emissions of the SN, but instead by shock heating the dense circumstellar medium of the AGB star progenitor. Thus, under this hypothesis we would expect to find a local stellar population with few or no massive stars ($\gtrsim 10M_{\odot}$).

More recently, \cite{Omand2024PulsarDriven} explored an alternate theory for the origin of SN 1054's peak luminosity. They fit a pulsar-driven supernova model to the historical observations of SN 1054’s luminosity. Their model suggests an initial spin-down luminosity of the Crab pulsar of around $10^{43-45}$erg/s with a spin-down timescale of 1-100 days and a low supernova explosion energy of  $\sim 10^{49}-10^{50}$ erg \citep{Omand2024PulsarDriven}. This implies a high initial pulsar rotational energy of $\sim 10^{50}$ erg and an initial spin period of $\sim13$ ms (other estimates of the initial spin periods are 15-20 ms, \citet{Kou2015}, and 3-5 ms, \citet{Atoyan1999}). They propose that the supernova underwent a "blowout", where the pulsar wind nebula broke through the ejecta shell, leaving dense filaments behind while accelerating the outer ejecta 50-200 years after the explosion.

Here we apply the \cite{velapulsar2022} approach for Vela to the Crab. In Section~\ref{sec:stellarpop} we describe the selection of stars surrounding the Crab pulsar and the spectral energy distributions (SED) of the most luminous stars. In Section~\ref{sec:analysis} we analyze the age distribution of the stars to estimate the likely mass of the Crab's progenitor. In Section~\ref{sec:results} we discuss the results. In Section ~\ref{sec:discussion/conclusions} we summarize our findings and potential future applications. 

\section{The Surrounding Stellar Population} \label{sec:stellarpop} 

We select stars near the Crab Pulsar using Gaia DR3 \citep{GaiaCollab2016, GaiaCollab2021}. Each star is required to have a parallax and all three Gaia Magnitudes (G, $R_{p}$ and $B_{p}$ bands). We use the position (J2000 05:34:31.947, +22:00:52.153) of the Crab pulsar from \cite{GaiaCollab2021} as the center. We use a weighted average parallax of $\varpi$ = $0.523\pm0.048$ mas ($d=1.912^{+ 0.43}_{-0.30}$ kpc)  for the Crab, combining the Gaia \citep[$\varpi$ = $0.511\pm0.078$ mas,][]{GaiaCollab2021}, and VLBI \citep[$\varpi$ = $0.523\pm0.048$ mas,][]{LinRebeccaVLBI2023} parallaxes. We first search for stars in a region centered on the position of the Crab with a maximum search angular size, $\theta = \sin^{-1}(R/D) = 3.58\degree$, where D = 2 kpc and $R = 125$ pc, with parallaxes $0 < \varpi < 1$ mas. We use a magnitude limit of $G < 12$ to include stars with absolute magnitudes $M_{G}\leq 0$ mag and masses that are $M\gtrsim1 M_{\odot}$.

\begin{figure}
    \includegraphics[width=\linewidth]{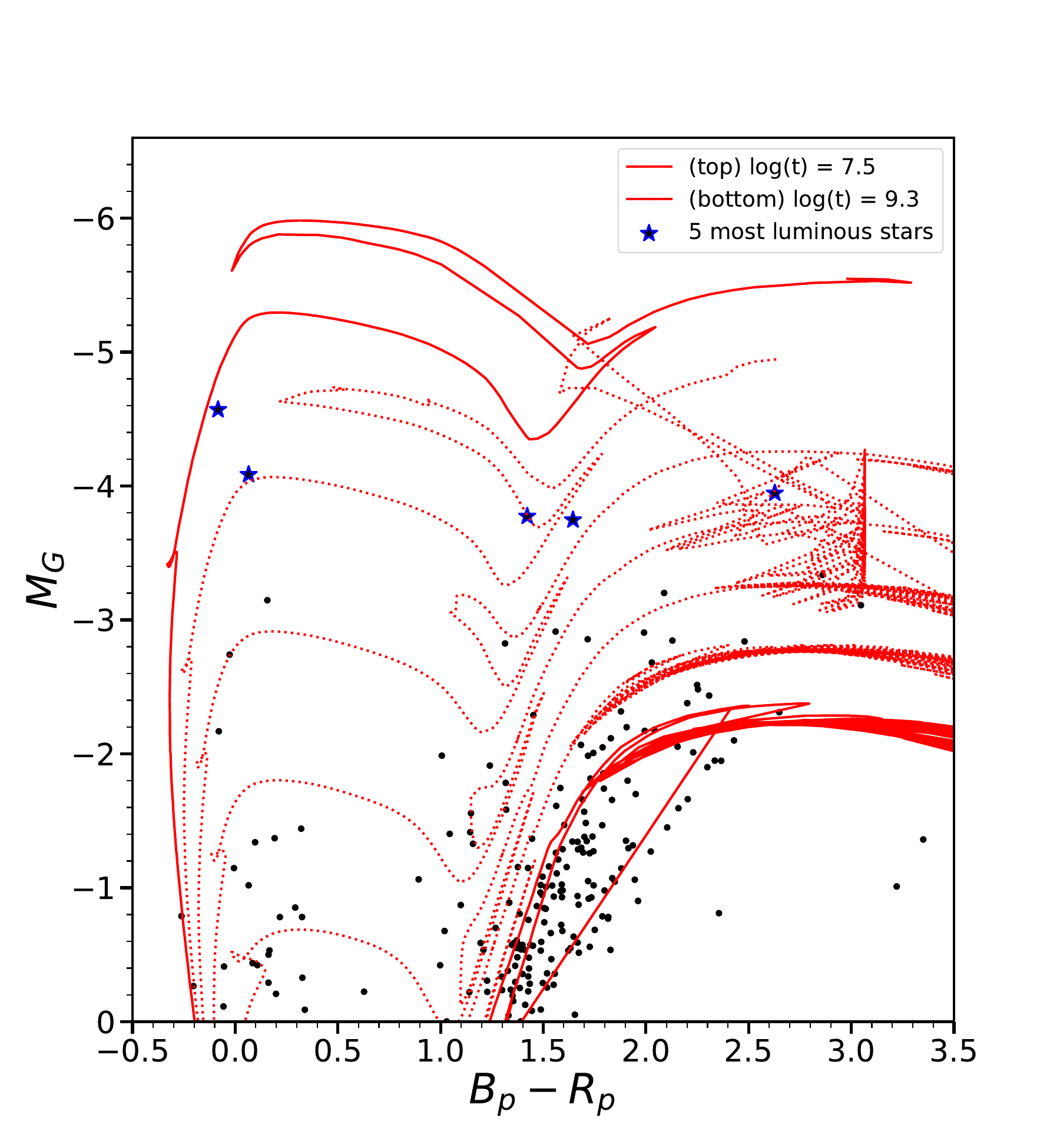}
    \caption{ Extinction corrected color-magnitude diagram of the stars local to the Crab from Gaia EDR3. Solar metallicity \texttt{PARSEC} isochrones are shown in red in age steps of 0.3 dex (top, solid: $\log_{10}(t) = 7.5$; bottom, solid: $\log_{10}(t) = 9.3$). The blue stars are the 5 most luminous stars for which we did individual SED fits.} 
    \label{fig:CMD_5}
\end{figure}

Geometrically the search region is a truncated cone and contains $1,525$ stars. Next, we convert the search region from a truncated cone into a truncated cylinder, with a radius $R = 100$ pc around the Crab. From the coordinates and parallax we form a vector ($\Vec{u}_{\star}$) for the position of the star. The cross product ($d_{\perp} = |\Vec{u}_{cp} \times \Vec{u}_{\star}|$) with a unit vector pointing to the pulsar ($\Vec{u}_{cp}$) provides the separation perpendicular to the line of sight and we keep the 225 stars with $d_{\perp} < R$ and parallaxes between $0.427 < \varpi < 0.619$ mas, which is the $2\sigma$ error range of the Crab's weighted average parallax. The resulting length of the cylinder ($\sim700$ pc) is longer than desired, but it seemed better to be "inclusive" given the distance uncertainties. We used extinction estimates for each star from the 3-dimensional (3D) \texttt{combined19 mwdust} models \citep{Bovy2016} which are based on \cite{Green2019} for the position of the Crab to obtain extinction corrected colors ($B_{p} - R_{p}$) and absolute magnitudes ($M_{G}$). We keep the $205$ stars with $-8 < M_{G} < 0$, $-0.5 < B_{p} - R_{p} < 3.5$. In practice, there are no $M_{G} < -8$ mag stars in the sample. In Figure \ref{fig:CMD_5} we show the CMD of these stars with the 5 most luminous stars labeled with blue stars, and \texttt{PARSEC} \citep[e.g.,][]{PARSEC2012Bressan, PARSEC2013MarigoAGB, PARSEC2020PastorelliTPAGB} isochrones spanning $10^{7.5}$ to $10^{9.3}$ years in steps of 0.3 dex.

\begin{table*}
    \caption{The luminous ($M_{G} < -3.5$) stars near the Crab Pulsar.}
    \label{tab:lumstars}
    \centering
    \renewcommand{\arraystretch}{1.5}
        \begin{adjustbox}{width=\textwidth}
            {\HUGE
            \begin{tabular}{c c c c c c c c}
            \toprule
            Star&$\chi^{2} / N_{dof} $&$\log(T_{\ast}) [K]$&$\log(L_{\ast}) [\lsun]$&$M_{\ast} [\msun]$&$\log(t) [yr]$&Sep [pc]&Comments \\ [0.5ex]
            \toprule
            HD 36879  & $ 1.94 $ & $ 4.205 \pm 0.018 $ & $4.381 \pm 0.063 $ & $ 10.74-11.66 $  & $ 7.27-7.33$ & $ 21.16 $ & O7V C \\ [0.5ex]
            
            HD 243780 & $ 3.86 $   & $ 3.643 \pm 0.010$ & $ 3.658 \pm 0.026 $& $ 6.38-8.26 $  & $ 7.54-7.81 $  & $55.84 $ & B0 E \\ [0.5ex]
            
            HD 36547  & $ 2.18 $ &  $4.283 \pm 0.009$ & $ 4.499 \pm 0.037 $ & $ 11.99-12.54 $  & $ 7.22-7.25 $  & $ 53.18 $ & B1III C \\ [0.5ex]
            
            IRAS 05310+2411& $ 2.34 $  & $ 3.598 \pm 0.007 $ & $ 3.512 \pm 0.017 $ & $ 4.18-5.23$  & $ 7.97-8.26 $  & $ 70.77 $ & No spectral type \\ [0.5ex]
            
            IRAS 05361+2406& $4.53 $   & $3.564 \pm 0.009$   & $3.809 \pm 0.025$  & $ 2.21-4.92 $  & $ 8.08-9.05 $  & $ 93.38 $ & Long Period Variable, No spectral type \\
            \hline
            \end{tabular}
            }
        \end{adjustbox}
\end{table*}

We first focus on these 5 most luminous stars $(M_{G} < -3.5)$ and fit their spectral energy distributions (SEDs) to estimate luminosities, temperatures and extinctions. We limit the SED fits to the most luminous stars because we are interested in the most massive and youngest stars local to the Crab. We use DUSTY \citep{ElitzuretalDUSTY2001} inside a Markov Chain Monte Carlo (MCMC) driver to optimize the SED fits and their uncertainties following methods of \cite{Adamsetal2017a} and \cite{velapulsar2022}. For the coolest stars we use MARCS \citep{MARCSGustafsson2008} stellar model atmospheres and \cite{CastelliandKurucz2003} otherwise. We used UV, optical, near-IR and mid-IR magnitudes from \cite{ThompsonetalUV1978} or \cite{WesseliusetalUV1982}, \cite{Johnsonetal1966} and ATLAS-REFCAT \citep{ATLASREFCAT2-2018ApJ}, 2MASS \citep{Cutrietal2MASS2003} and ALLWISE \citep{CutriWISE2014}. We use temperature and extinction priors based on spectral types reported in VizieR \citep{vizier2000} and the widths of the temperature and extinction prior errors were $\pm1000 K$ and $\pm0.1$ mag.


\begin{figure}
    \includegraphics[width=\linewidth]{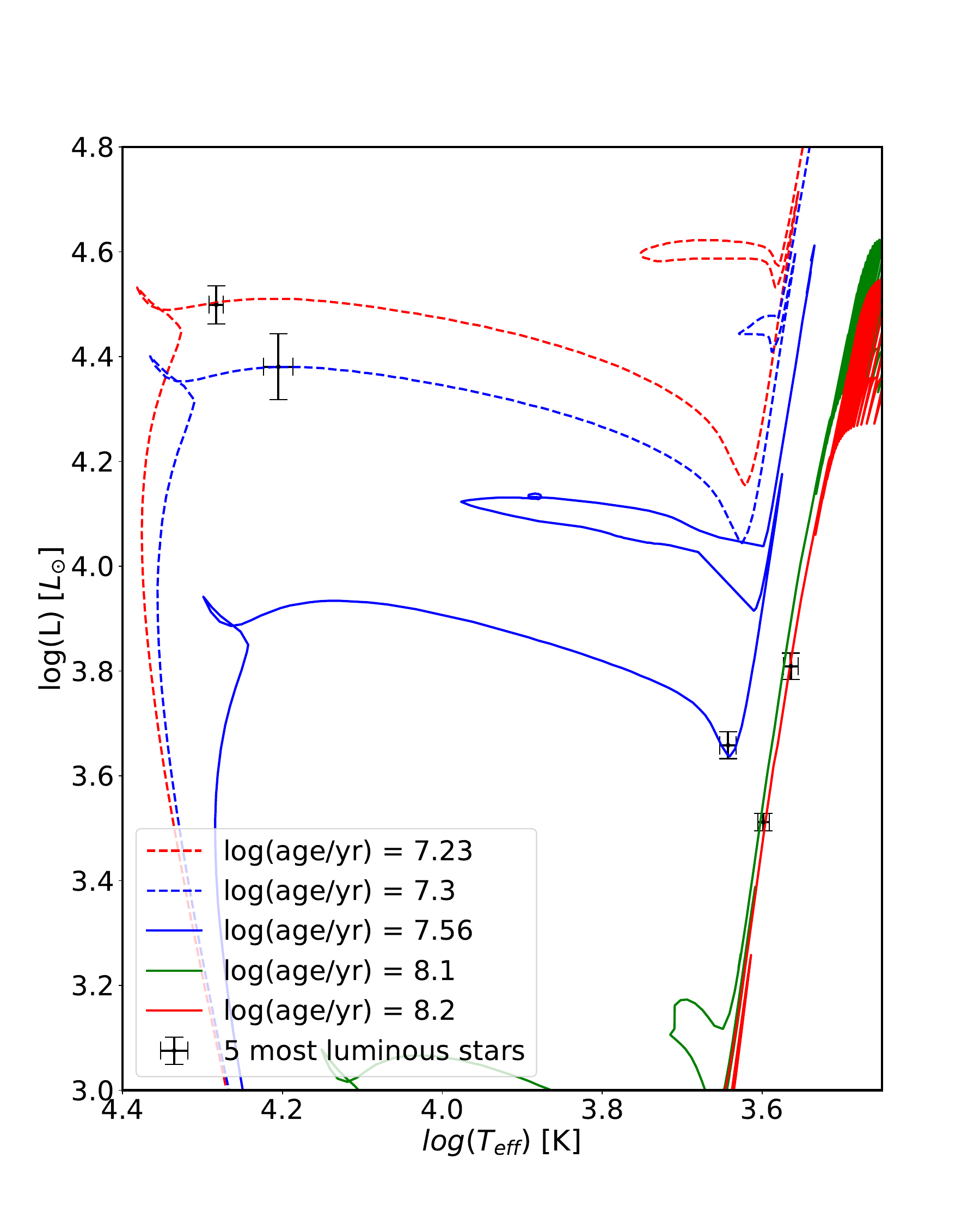}
    \caption{Estimated luminosities and temperatures of the 5 most luminous stars obtained from SED fits as compared to \texttt{PARSEC} Solar metallicity isochrones with ages from $10^{6.3}yr$ (dashed red) to $10^{10.1}yr$ (dashed grey) in steps of 0.01 dex.}
    \label{fig:SEDfit_15results}
\end{figure}

Table ~\ref{tab:lumstars} provides the estimated age, mass range, luminosity, temperature, known or unknown spectral classification, transverse distance from the Crab pulsar, and the goodness of the SED fits. The age and mass constraints were obtained by finding the \texttt{PARSEC} isochrone ages and masses where the luminosity and temperature are within the 1-sigma error range of the SED fits. We use Solar metallicity \texttt{PARSEC} isochrones with ages from $10^{6.3}$ yr to $10^{10.1}$ yr in steps of 0.01 dex. The five most luminous stars have luminosities of  $10^{4.38\pm0.06} L_{\odot}$,  $10^{3.65\pm0.03} L_{\odot}$,  $10^{4.49\pm0.04} L_{\odot}$, $10^{3.51\pm0.02} L_{\odot}$, $10^{3.81\pm0.03} L_{\odot}$, with masses of $11.19M_{\odot}\pm0.07M_{\odot}$, $7.32M_{\odot}\pm0.15M_{\odot}$, $12.33M_{\odot}\pm0.09M_{\odot}$, $4.75M_{\odot}\pm0.06M_{\odot}$, $4.34M_{\odot}\pm0.08M_{\odot}$, respectively. The two most massive stars both lie near the main sequence and have luminous/early spectral types (O7 and B1). The other $3$ stars are much less massive red giants. The spectral type reported for HD243780 (B0 E) is inconsistent with its location on the CMD and the SED model. We show the SED models in Appendix \ref{appendix:A}. Figure \ref{fig:SEDfit_15results} shows the resulting SED fit temperatures and luminosities of the 5 most luminous stars on a Hertzsprung-Russell diagram.
\section{Progenitor Mass Analysis}
\label{sec:analysis}

The next step in estimating the progenitor mass of the Crab is to find the age distribution of the selected stars. We use 13 age bins ($i = 1...13$) with 0.3 dex widths. We assume single star evolution and Solar metallicity. We randomly draw $N_{trial} = 3\times10^{8}$ stars from a Salpeter IMF with a minimum mass of $M_{min} = 1 M_{\odot}$. We obtain their color and magnitudes using the \texttt{PARSEC} isochrone models sampled with $\Delta \log t = 0.01$ dex \citep{PARSEC2012Bressan, PARSEC2013MarigoAGB, PARSEC2020PastorelliTPAGB}. We create the density maps $F_{jk}(t_{i}) = F_{jk}^{i}$ of the stars in a Gaia CMD where $i$, $j$, and $k$ index the time, absolute magnitude, and color, respectively.  Each modeled star is created by uniformly selecting a time between $t_{min,i}$ and $t_{max,i}$, corresponding to a constant star formation rate for each age bin. A star is added to the density map if the chosen mass still exists on the isochrone and the star lies in the absolute magnitude range of $0.0 > M_{G} > -8.0 $ (index $j$) and the color range $-0.5 < B_{p} - R_{p} < 3.5 $ (index $k$). For each star falling within these ranges of color and absolute magnitude, we add a $1$ to the cell $[j,k]$ corresponding to their color and magnitude. We do not include either observed or model stars that are either too red, too blue, or too luminous. We test the effects of this selection by creating maps that have these stars added to the edges of the density distributions, but this had no effect on the results (see Section \ref{sec:results}).


We apply the \texttt{mwdust} extinction corrections to the observed stars surrounding the Crab and compare the stars to the density maps using their extinction corrected photometry (Fig. \ref{fig:CMD_5}). We assume that these \texttt{mwdust} estimates are correct on average. We examined the effects of extinction uncertainties on the model density maps by also producing density maps with random Gaussian extinctions of  $\sigma_{E(B-V)} = 0.01$, $0.03$ and $0.1$ mag added to each trial star. 

For a constant star formation rate (SFR), the formation rate by mass for $M > M_{min}$ is 
\begin{equation} \label{eq:1}
    \frac{dN}{dMdt} = \frac{(x-2) SFR}{M^{2}_{min}} \bigg(\frac{M}{M_{min}}\bigg)^{-x}
\end{equation}

\noindent with x = 2.35 and a mean mass of $\langle M \rangle = (x-1) M_{min}/(x-2)$. The age bins are in logarithmic time intervals, $t_{min,i} < t < t_{max,i}$, where $\Delta t = t_{max,i}-t_{min,i}$. Since the $SFR_{i}$ is constant, the number of $M > M_{min}$ stars formed in each interval is $N_{i} = SFR_{i} \Delta t_{i} / \langle M \rangle $. The number of stars that die in a short time interval $\delta t$ today is 
\begin{equation} \label{eq:2}
    N_{i} \frac{\delta t}{\Delta t_{i}} \biggr[ \bigg(\frac{M(t_{min,i})}{M_{min}}\bigg)^{(1-x)} - \bigg(\frac{M(t_{max,i})}{M_{min}}\bigg)^{(1-x)} \biggr] = N_{i}S_{i}\delta t 
\end{equation}

\noindent where M(t) is the most massive surviving star on the isochrone, and $S_{i}\delta t$ is the fraction of $M>M_{min}$ stars that died in the last $\delta t$ years. A full derivation of Equation \ref{eq:2} is in Appendix \ref{appendix:B}.




The observed stars can be placed on the absolute magnitude and color grid in the same manner, with $N_{jk}^{\star}$ stars in a pixel and $\sum_{jk}N_{jk}^{\star} = N^{\star} = 205$. The number of model stars in a given magnitude and color bin is $N_{jk} = \sum_{i} \alpha_{i} F_{jk}^{i}$, where $\alpha_{i}$ is proportional to the star formation rate of age bin $i$, and the model has a total of $N = \sum_{jk} N_{jk}$ stars. The Poisson probability of finding the observed number of stars in a bin of color and magnitude is





\begin{equation} \label{eq:3}
    \frac{N_{jk}^{N_{jk}^{\star}} e^{-N_{jk}}}{N_{jk}^{\star}!} ,
\end{equation}

\noindent so the logarithm of the likelihood  for all $N^{\star}$ stars is 

\begin{equation} \label{eq:4}
    \ln L = \sum_{jk} \ln \bigg({r N_{jk}^{N_{jk}^{\star}}} \bigg) - \sum_{jk} r N_{jk} ,
\end{equation}

\noindent where the first term is the sum over bins containing stars and the second is the sum over all bins. We discard the factorial $N^{\star}_{jk}!$ because the calculation depends only on likelihood differences and not the absolute likelihood. Empty cells of $N_{jk}$ are filled with a small number ($1\times 10^{-32}$) to avoid numerical problems.

We introduced a "re-normalization" factor $r$ in Equation \ref{eq:5}. Equation \ref{eq:5} with $r=1$ will include Poisson fluctuations in N relative to $N^{\star}$. However, we really want the probability for how the $N^{\star}$ stars are divided over the $13$ age bins. If we choose 


\begin{equation} \label{eq:5}
    r = N_{\star} \bigg[ \sum_{jk} \sum_{i} \alpha_{i} F_{jk}^{i} \bigg]^{-1} ,
\end{equation}

\noindent and then re-normalize $\alpha_{i} \rightarrow r \alpha_{i}$ so that $\sum_{jk} \sum_{i} \alpha_{i} F_{jk}^{i} \equiv N_{\star}$, the likelihood becomes the multinomial likelihood for how to divide the $N_{\star}$ stars over the age bins. Some age bins are susceptible to $\log(N_{i}) \rightarrow - \infty $. To prevent such numerical divergences, we add a weak prior of 
\begin{equation} \label{eq:6}
     - \lambda^{-2}\sum_{i}\bigg[\ln\bigg(\frac{\alpha_{i}\Delta t_{i+1}}{\alpha_{i+1}\Delta t_{i}}\bigg)\bigg]^{2} -\lambda^{-1}\sum_{i}\bigg[\ln\bigg(\frac{\alpha_{i}}{\alpha_{0}}\bigg)\bigg]^{2}  , 
\end{equation}
\noindent where $\lambda = 6.91$, which adds a penalty of unity to the likelihood if adjacent bins have star formation rates ($SFR_{i} \approx N_{i}/\Delta t_{i}$) that differ by a factor of 1000, and the $2^{nd}$ term with $\alpha_{0} = N_{\star} / N$ penalizes not distributing the observed stars uniformly over the age bins.  

We calculate the number of deaths in a time range $\delta t$ with $N_{i}S_{i}\delta t$, where $S_{i}$ is independent of $\delta t$. We want to find the probability that the progenitor was born from age bin i versus j, and therefore use the normalized probability for each age bin of

\begin{equation}
    \frac{P_{i}}{P_{tot}} = \frac{N_{i}S_{i}}{\sum_{all}N_{i}S_{i}} ,
\end{equation}

\noindent is independent of $\delta t$, which has a total probability of unity. 

We optimize the likelihood and estimate the uncertainties using the Monte Carlo Markov Chain (MCMC) driver in the \texttt{emcee} python package \citep{Foreman-Mackey2013emceeMCMC}, with $\log \alpha_{i}$ as the fit parameters. We uses $300$ walkers each with a chain length $10,000$. We discard the first $1000$ entries of each walker chain for determining uncertainties. Within the MCMC driver, these $\log \alpha_{i}$ are re-normalized (Eq. \ref{eq:5}) before calculating the likelihood.

\section{Results}
\label{sec:results}
\begin{figure}
    \includegraphics[width=\linewidth]{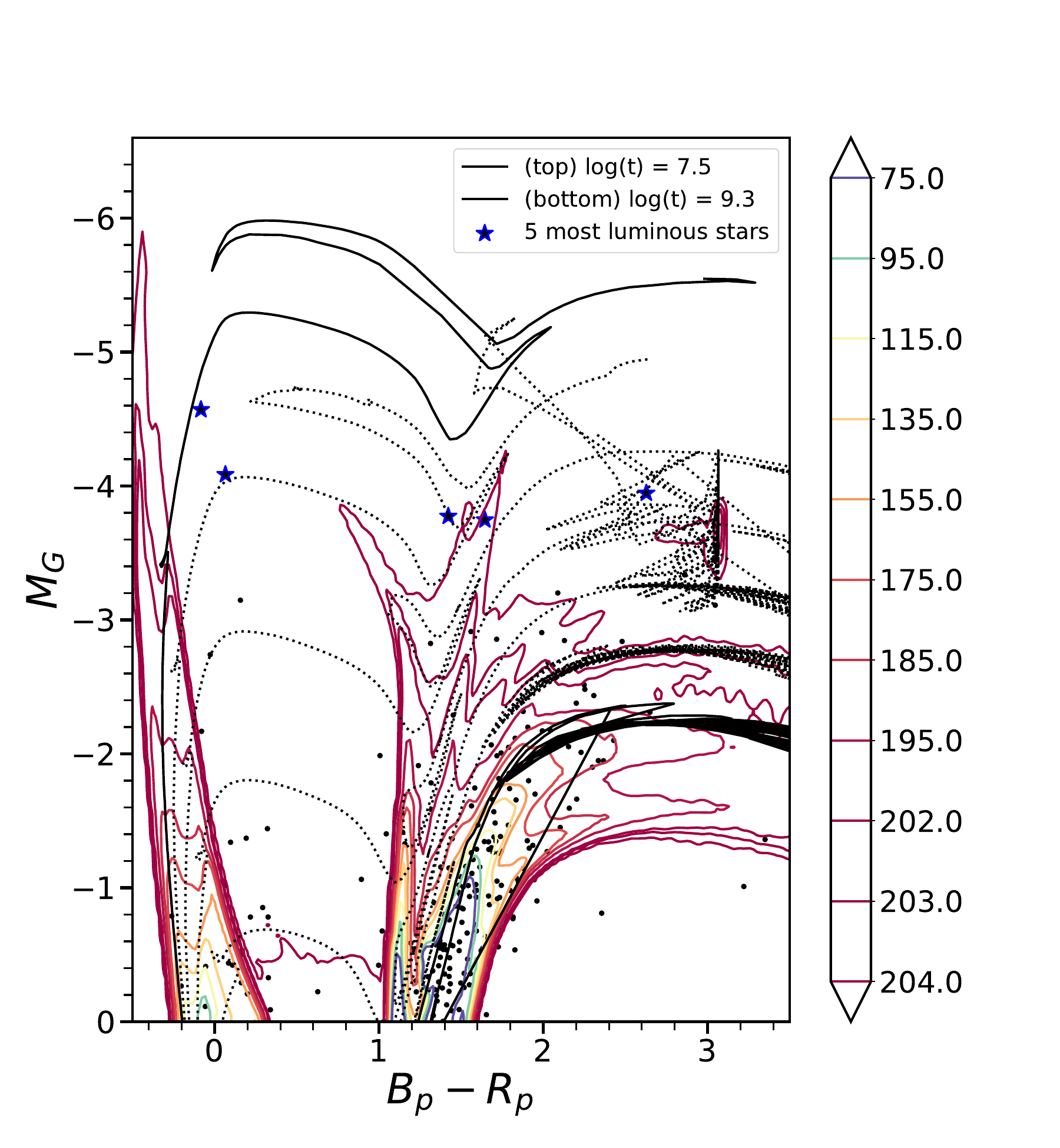}
    \caption{The black dots are the extinction corrected Gaia CMD of the stars near the Crab, and the curves are the Solar metallicity \texttt{PARSEC} isochrones with ages from $10^{7.5}$yr (top black) to $10^{9.3}$yr (bottom black) in steps of 0.3 dex. The model density contours are drawn at the level which encompasses the number of stars shown on the scale bar. The blue stars are the 5 most luminous stars near the Crab.} 
    \label{fig:crab_stellar_density_contour_13bins}
\end{figure}

Figure \ref{fig:crab_stellar_density_contour_13bins} shows the resulting density contours for the distribution of the model stars in the CMD in the maximum likelihood model. The purple (magenta) contour lines show the low (high) stellar densities. The maximum densities lie along the main sequence and the red giant branch as expected and largely encompasses the observed stars. 



\begin{figure}
    \includegraphics[width=\linewidth]{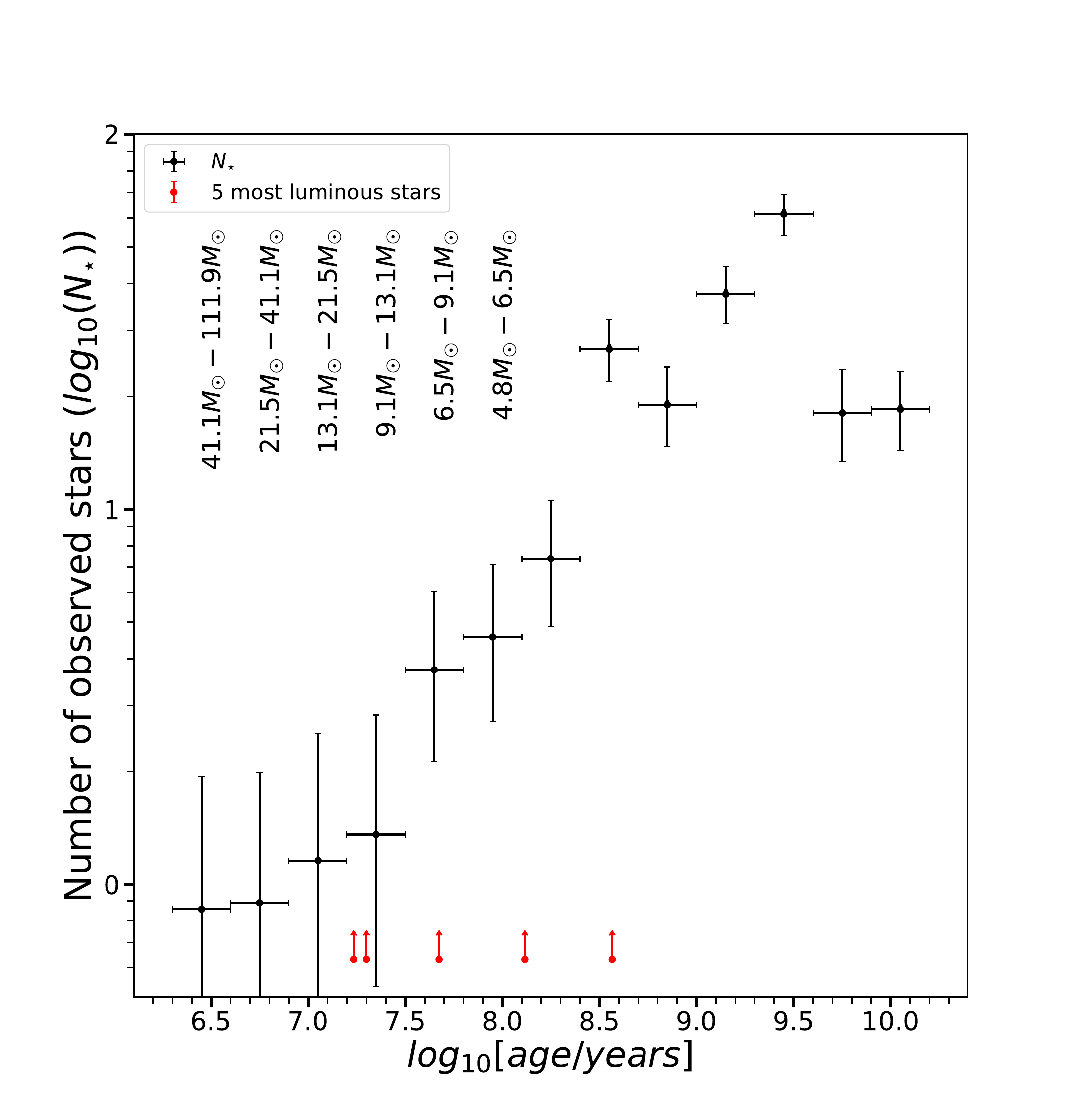}
    \caption{The number of observed stars assigned to each age bin (black points), where the horizontal bar shows the width of the bin in age. The mass range for each age bin is listed at the top left of the plot. The points and vertical errorbars are the median and 16 and 84 percentile ranges of the number of stars associates with each age bin. The red arrows show the estimated ages of the 5 individually modeled stars.}
    \label{fig:number_vs_agebin_13bins}
\end{figure}


\begin{figure}
    \includegraphics[width=\linewidth]{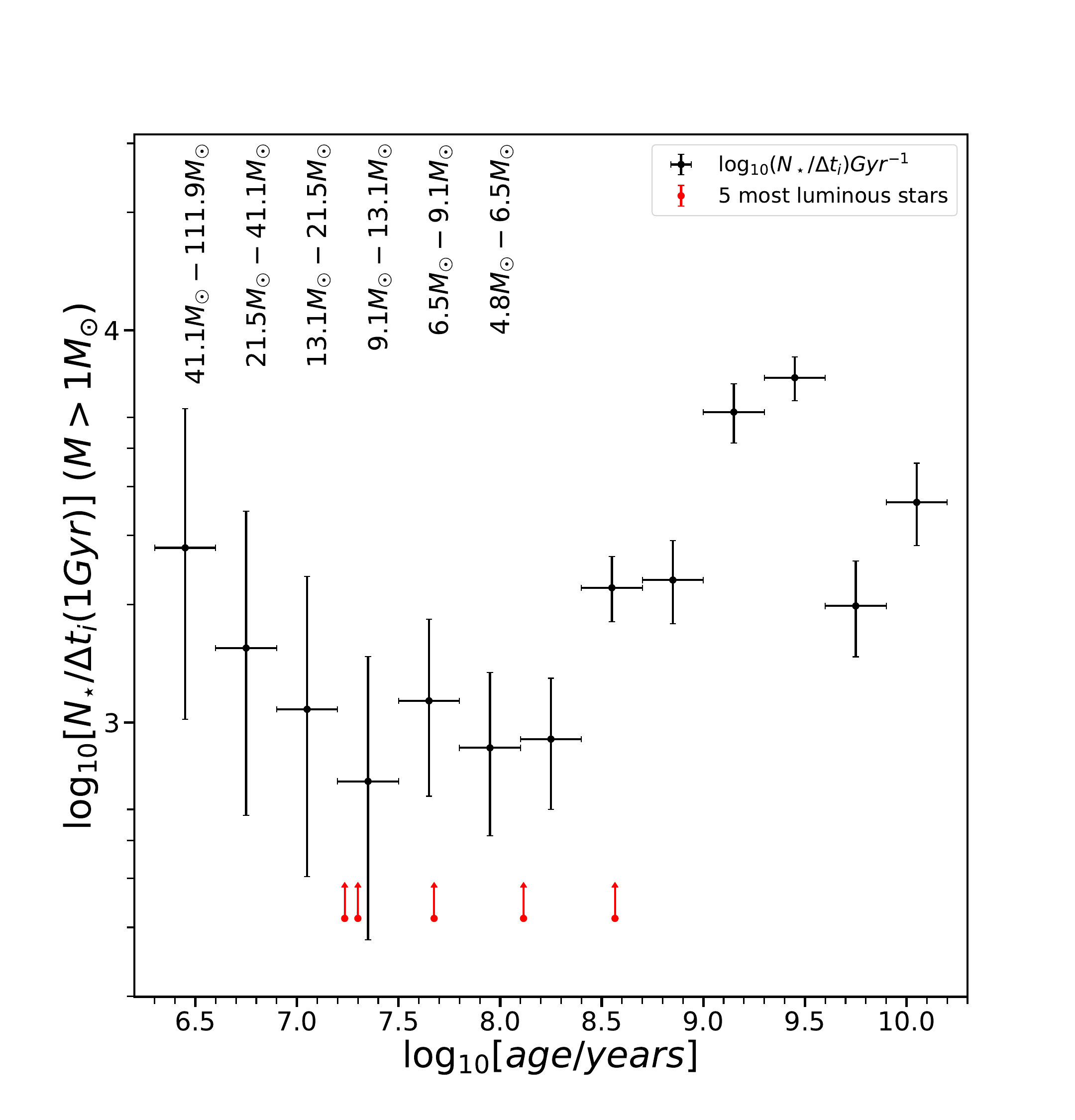}
    \caption{The number $N_{\star,i}$ of $M > 1 M_{\odot}$ stars formed in each age bin per $10^9$ years. This corresponds  to the number of observed stars shown in Figure \ref{fig:number_vs_agebin_13bins} divided by the fraction of the Monte Carlo trials leading to a star on the density grid and the temporal age bin width, $\Delta t_i$ (in units of $1 Gyr = 10^9 yr$). The horizontal errors are the 0.3 dex widths of the age bins. The red arrows are the estimated ages of the 5 most luminous stars.}
    \label{fig:N_i_vs_agebin_13bins}
\end{figure}

\begin{figure}
    \includegraphics[width=\linewidth]{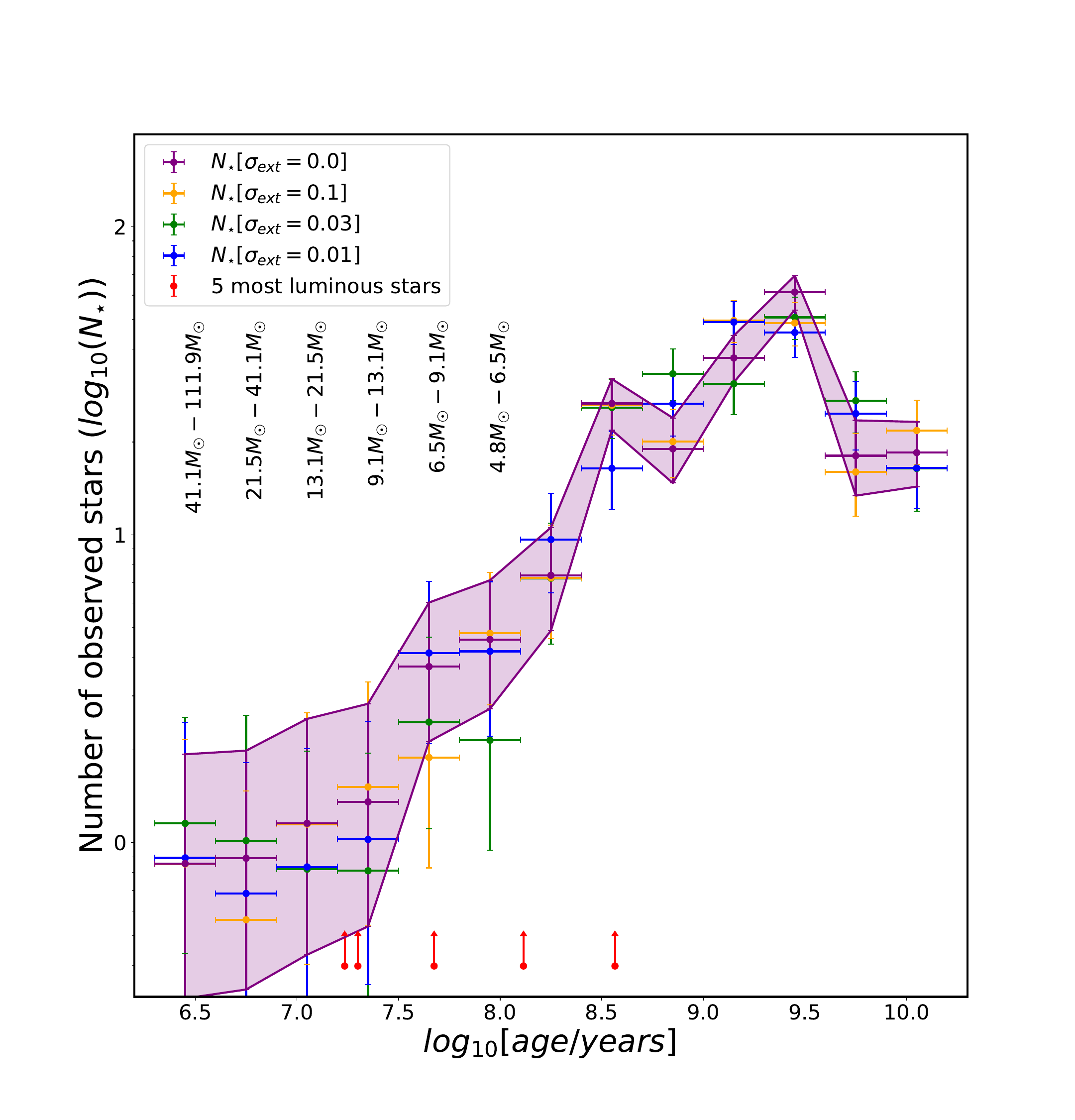}
    \caption{The model distributions of the observed stars in the age bins using density grids with an extinction scatter of $\sigma_{ext} = 0.1, 0.03, 0.01$ (orange, green, and blue, respectively). The number of observed stars in Fig. \ref{fig:number_vs_agebin_13bins} are plotted in purple, $\sigma_{ext} = 0.0$. The red arrows show the estimated ages of the 5 most luminous stars.}
    \label{fig:compiled_gaus_all_number_vs_agebin_nosidebins}
\end{figure}

\begin{figure}
    \includegraphics[width=\linewidth]{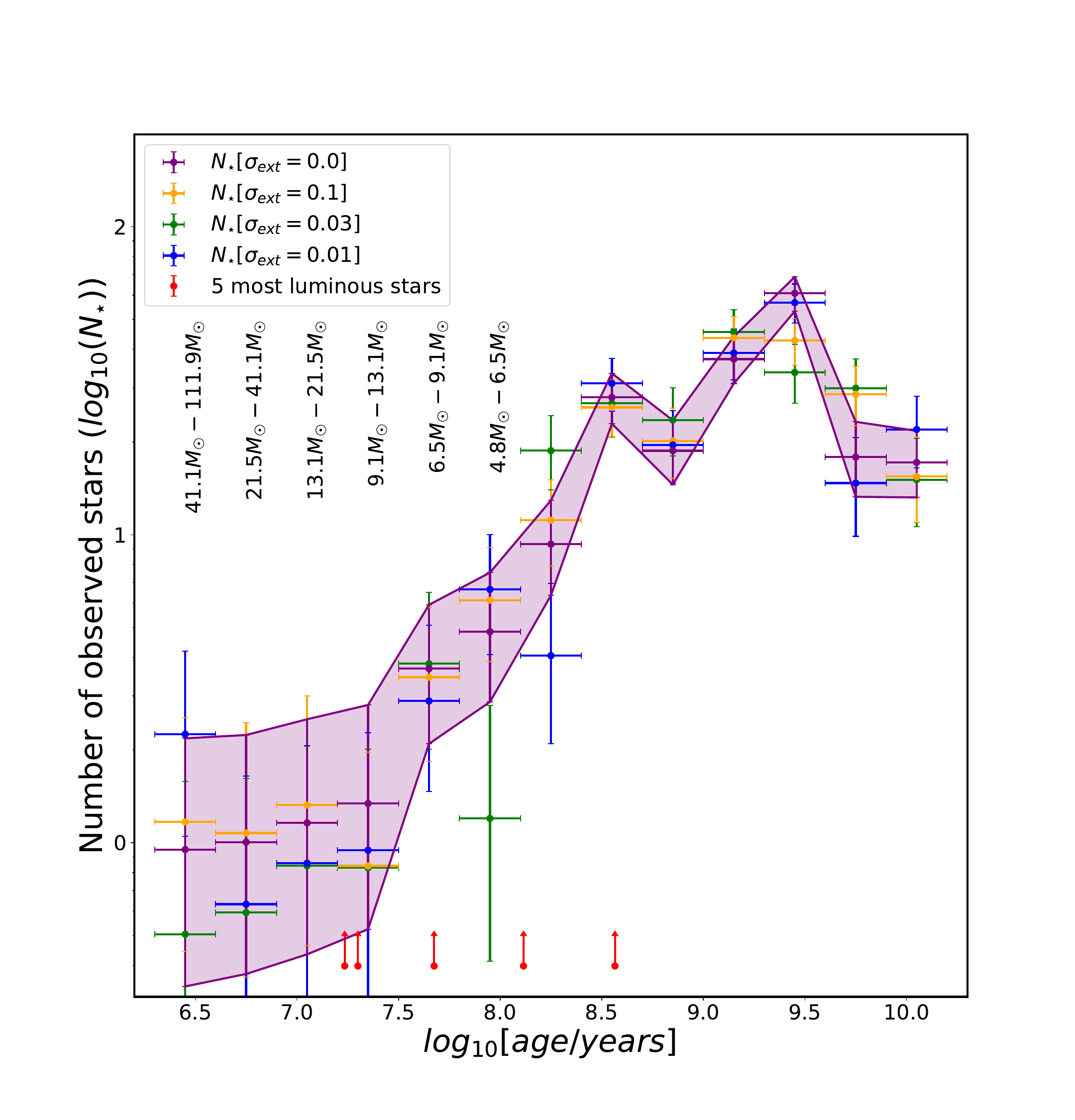}
    \caption{The model distribution of the observed stars across the age bins using modeled density maps with the alternate treatment of edge effects and $\sigma_{ext} = 0.1, 0.03, 0.01$, shown in orange, green, and blue, respectively. The vertical error bars span the $16^{th}$ and $84^{th}$ percentiles. The number of observed stars with the standard treatment of edge effects and no extinction scatter ($\sigma_{ext} = 0.0$) are plotted in purple. The red arrows are the estimated ages of the 5 most luminous stars.}
    \label{fig:compiled_gaus_all_number_vs_agebin_sidebins}
\end{figure}


Figures \ref{fig:number_vs_agebin_13bins} and  \ref{fig:N_i_vs_agebin_13bins} show two different ways of viewing the distribution of stars in age. Figure \ref{fig:number_vs_agebin_13bins} shows the distribution of the $N_{\star} = 205$ modeled stars over the age bins. The total number of stars exactly equals to 205 because of the renormalization in Equation \ref{eq:5}. The fourth youngest age bin is $10^{7.2}yr - 10^{7.5}yr$, which corresponds to mass ranges of $9.1 M_{\odot} - 13.1 M_{\odot}$, contains 2 stars and the following age bin $10^{7.5}yr - 10^{7.8} yr$ ($6.5 M_{\odot} - 9.1 M_{\odot}$) contains 4 stars. Consistent with Figure \ref{fig:crab_stellar_density_contour_13bins}, there are very few high mass stars in the region local to the Crab. The next two age bins corresponding to ages $10^{7.8}yr -10^{8.4} yr$ ($3.7 M_{\odot} - 6.5 M_{\odot}$) have a cumulative of 13 stars. The eighth oldest age bin ($10^{8.4}yr -10^{8.7} yr$) has 29 stars, which corresponds to $2.9 M_{\odot} - 3.7 M_{\odot}$ stars. Figure \ref{fig:number_vs_agebin_13bins} also shows the estimated ages of the 5 most luminous stars from the SED fits. The two youngest of these stars fall in the fourth youngest age bin ($10^{7.2}yr - 10^{7.5} yr$ or $9.1 M_{\odot} - 13.1 M_{\odot}$), which is also consistent with the 4 stars found in the model. Figure \ref{fig:N_i_vs_agebin_13bins} shows the number $N_{i}$ of $M > 1 M_{\odot}$ stars formed per $10^9$ years as a function of age. This is simply the observed number of stars (Fig. \ref{fig:number_vs_agebin_13bins}) divided by the fraction of the Monte Carlo trials leading to a star on the density grid and the temporal width of the bin.

Figure \ref{fig:compiled_gaus_all_number_vs_agebin_nosidebins} compares the result in Fig. \ref{fig:number_vs_agebin_13bins} to the result when we include a scatter in the extinction of $\sigma_{ext} = 0.1, 0.03, 0.01$ mag, and Fig. \ref{fig:compiled_gaus_all_number_vs_agebin_sidebins} does the same but also includes stars which are too red or too blue on the grid edges. The qualitative structure of Fig. \ref{fig:number_vs_agebin_13bins} is little changed by these variations.

\begin{figure}
    \includegraphics[width=\linewidth]{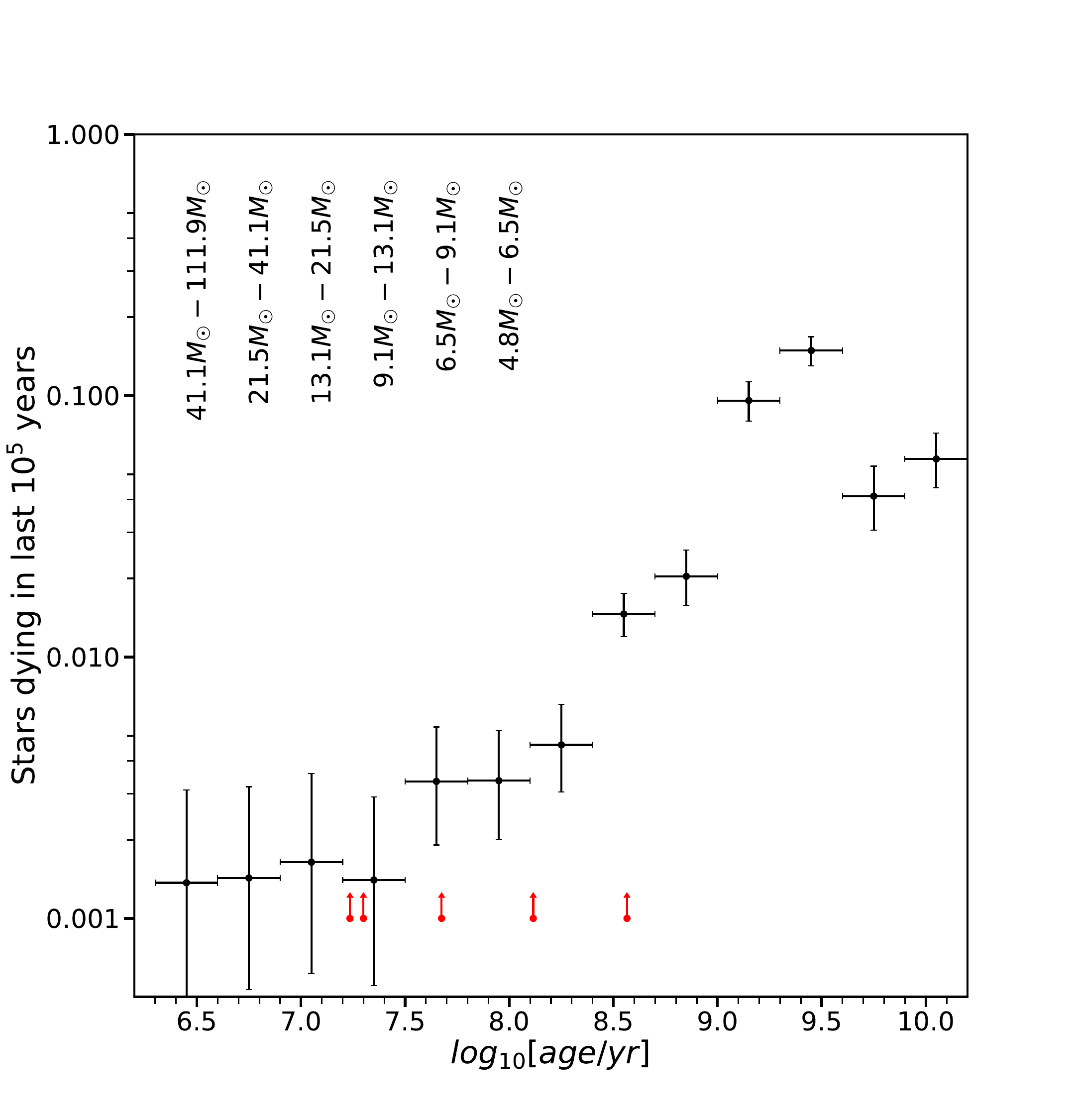}
    \caption{The number of stellar deaths over the last $10^{5}$ years. The points are the estimate probability for each bin and its 1$\sigma$ confidence range. The horizontal error bars span the 0.3 dex age bin widths. The arrows show the ages of the 5 most luminous stars. }
    \label{fig:N_dt_vs_agebin_13bins}
\end{figure}

\begin{figure}
    \includegraphics[width=\linewidth]{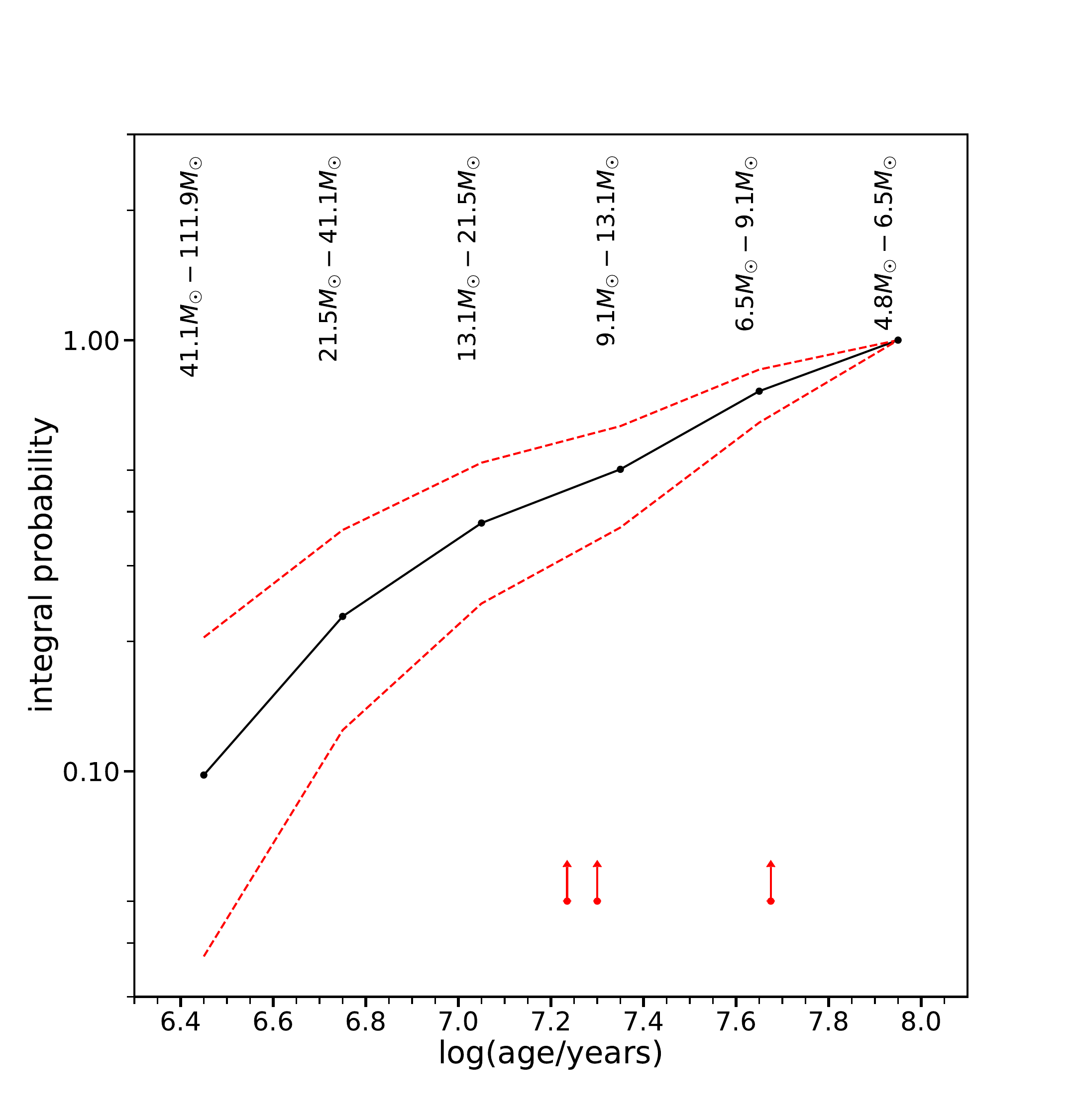}
    \caption{Integral probability distribution of the number of stellar deaths as a function of age. The red dashed lines are the $1\sigma$ confidence range for the median number of stellar deaths within each age bin. The arrows are the age estimates of the 3 most luminous stars. The distribution is truncated on the oldest age bin that can produce a ccSN, albeit through the explosion of a merger remnant.}
    \label{fig:integral_prob_13bins}
\end{figure}

Figures \ref{fig:N_dt_vs_agebin_13bins} and \ref{fig:integral_prob_13bins} show the differential and integral distributions in age of the number of stars expected to have died in the last $\delta t=10^5 yr$. The probabilities are low because if we took a random volume with this stellar age distribution, the probability of finding an SNR would be very low. By selecting the volume to contain an SNR, it is no longer random. This selection effect only affects the absolute probabilities and not the relative probabilities. Formally, we find that lower mass progenitors are favored, but the probability contrast between the lower and higher masses does not allow a very strong limit. If we focus on the  differential probability, the age bins corresponding to stars which might be electron capture supernovae either directly or as a binary merger product (the $10^{7.2}yr - 10^{7.8} yr$ age bins with a mass range of $6.5 M_{\odot} - 13.1 M_{\odot}$), they have median likelihoods roughly 5 times those of the higher mass ($\geq 13.1 M_{\odot}$) bins. However, if we consider the integral probability distribution over this same age range, these two bins encompass only $64\%$ of the probability. If we include the next lower age bin ($10^{7.8} yr - 10^{8.1} yr$ with masses $4.8 M_{\odot} - 6.5 M_{\odot}$), this increases to $76\%$ of the probability. Essentially, the dynamic range of the differential probability distribution simply is not large enough to strongly rule out higher mass progenitors. The absence of any luminous stars for age bins younger than $10^{7.2}$ years does suggest that these younger ages should be disfavored.

\section{Discussion}
\label{sec:discussion/conclusions}


    

We examine the properties of the 205 stars with $M_{G} < 0$ mag in a volume surrounding the Crab SNR. If we examine the five most luminous stars, we find that the two most luminous, HD 36879 and HD 36547, have luminosities, masses and ages of roughly $10^{4.38 \pm 0.06}L_{\odot}$ $(10^{4.49\pm 0.04}L_{\odot})$, $11.19M_{\odot} \pm 0.07M_{\odot} $ $(12.33 M_{\odot} \pm 0.09 M_{\odot})$, and $10^{7.30} yr$ ($10^{7.23} yr$). Both are main sequence or perhaps slightly evolved blue stars. If we analyze the overall age distribution of all these stars and estimate the likely age distribution of stars which will have recently died, we find modest evidence in favor of lower mass stars, consistent with the proposal that the progenitor was an extreme AGB star leading to an electron capture supernova. The age bin where the progenitor could be the explosion of a binary merger \citep{Zapartas2017A&A...601A..29Z} is roughly likely as the age bin corresponding to a directly formed AGB star. This is interesting since the Crab was not a binary at death \citep{NotBinariesatdeathKochanek2018}, but almost all massive stars start in binary or high order systems \citep[e.g.,][]{Moe&DiStegano2013closebinarymassivestars, Sanaetal2012binarymassivestars}. Unfortunately, the probability distribution does not drop sufficiently rapidly towards younger, higher mass progenitors to make a strong statistical case for this scenario. These results are stable with respect to the treatment of edge effects and allowing for noise in the stellar extinction estimates. One problem is that the parallax of the Crab is still sufficiently uncertain that we have had to use a distance range of roughly $\sim 700$ pc which is much larger than desirable. \cite{velapulsar2022}, in a similar analysis of the stellar populations around Vela, found that contamination was already likely significant for a distance range of only $\sim200$ pc. 

Nonetheless, this still seems to be an interesting new probe of Galactic SNRs given that we have so few. \cite{Kochanek2024S147} demonstrate a new method for estimating distances using multi-object Hectochelle spectrographs to search for the appearance of high velocity absorption features in stars behind the SNR that can provide distances to the typical SNR where there is no parallax for a remnant. Even in cases like the Crab with a parallax measurement, this approach may still do better than a direct parallax because it can average over the parallaxes of multiple stars rather than a single object. Some SNRs of particular interest are the ones which are presently interacting binaries \citep[SS 433, HESS J0632+057, 1FGL J1018.6-5856;][respectively]{SS433-distance-2004, Hinton-Hess2009ApJ, Corbetbinary2011, 1FGLJ10186-dists-Ackermann2012} and the one relatively clear case of a binary unbound in the supernova \citep[S147,][]{Dincel2015, unboundbinaryCK2021}.

\section*{Data Availability Statement}

All data used in this analysis are publicly available.

\section*{Acknowledgements}
Elvira Cruz-Cruz is supported by NASA FINESST Fellowship 80NSSC23K1444. Christopher S. Kochanek is supported by NSF grants AST-2307385 and AST-2407206. This research has made use of the VizieR catalogue access tool, CDS,
Strasbourg, France \citep{10.26093/cds/vizier}. 




\bibliographystyle{mnras}
\bibliography{bib} 



\newpage
\begin{appendix}
    
\section{SED fit results}
\label{appendix:A}
\begin{figure}
    \includegraphics[width=\linewidth]{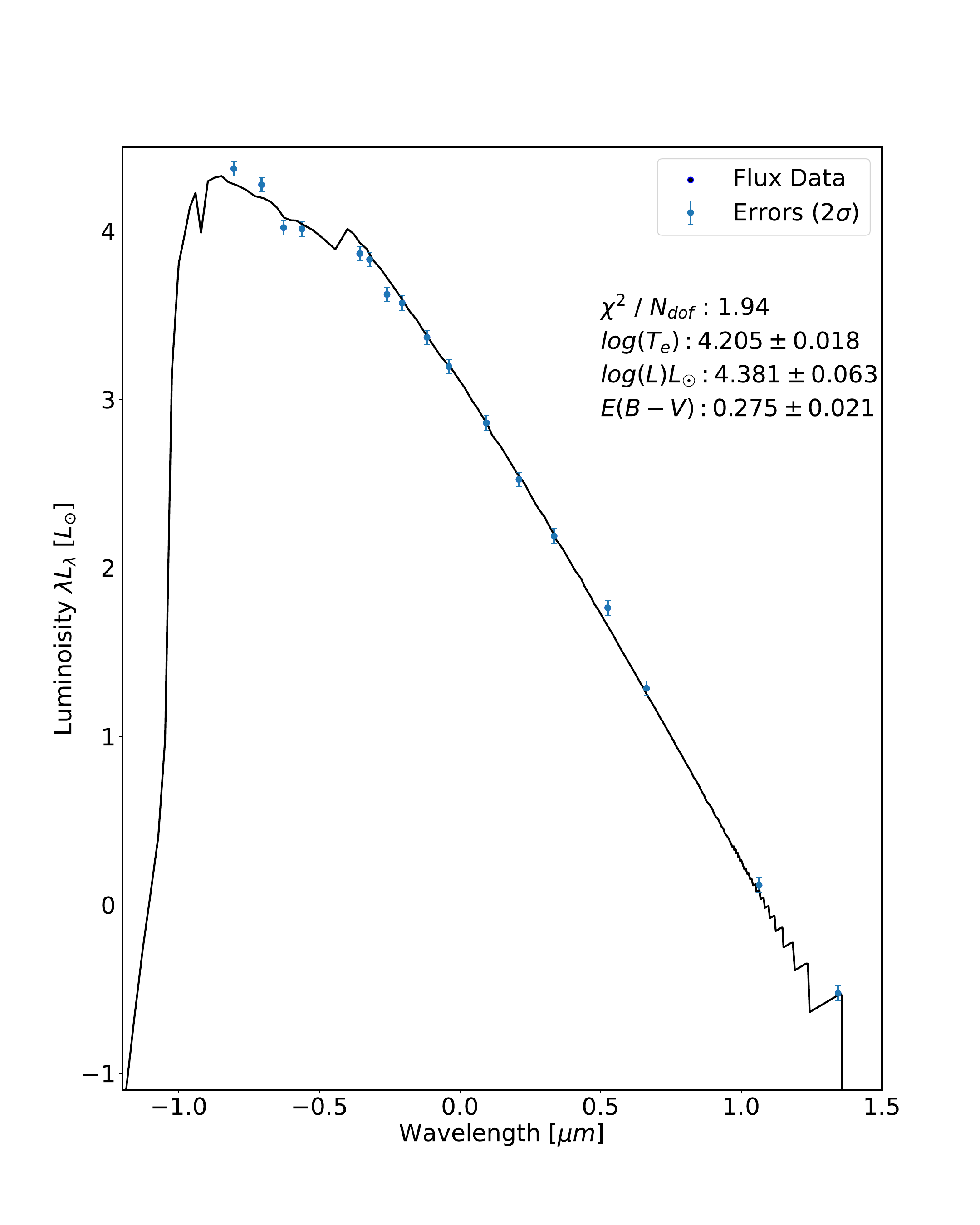}
    \caption{The SED of the O7V(n)(f)z C star HD 36879.}
    \label{fig:star_1}
\end{figure}

\begin{figure}
    \includegraphics[width=\linewidth]{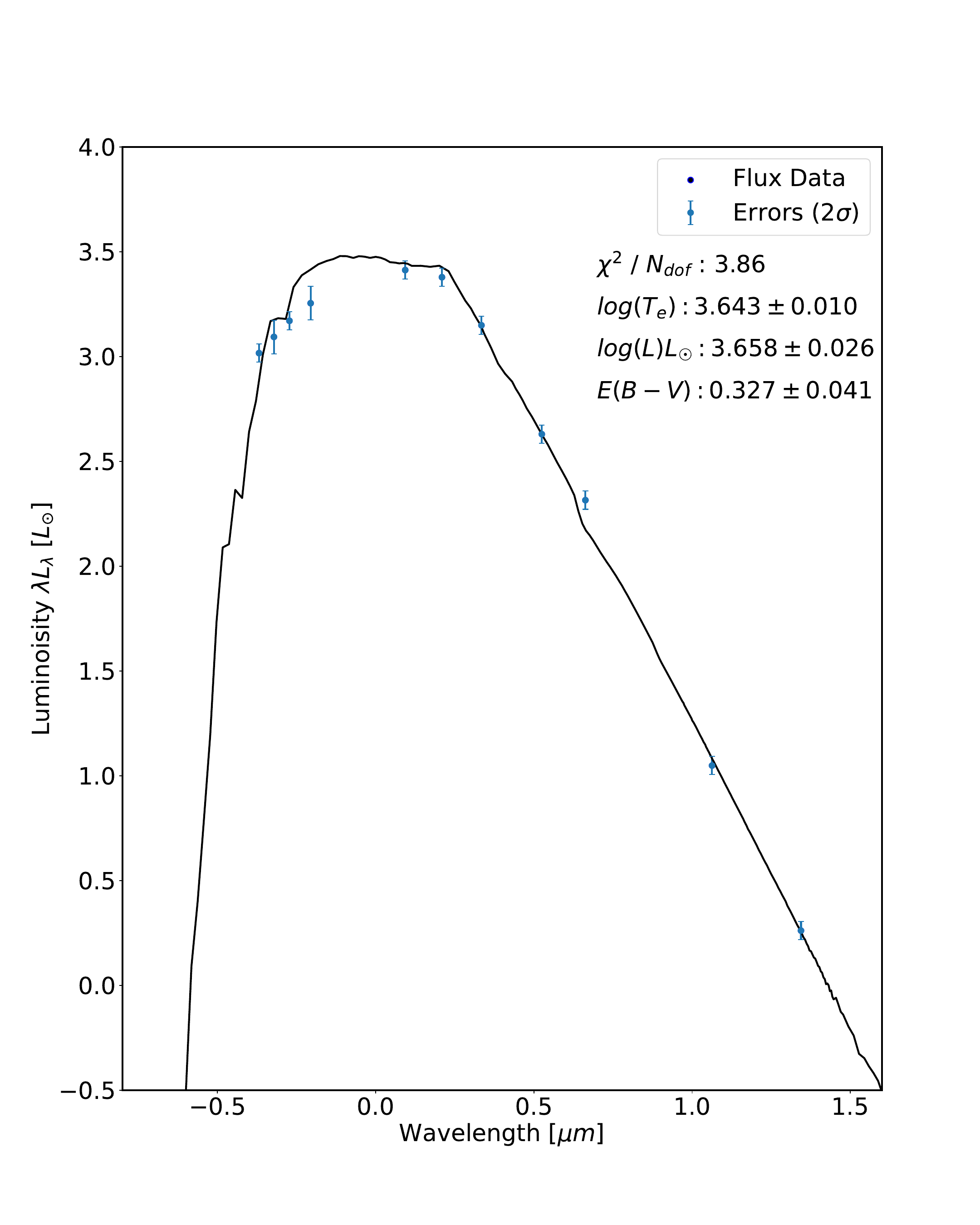}
    \caption{The SED of the B0 E Star HD 243780.}
    \label{fig:star_2}
\end{figure}

\begin{figure}
    \includegraphics[width=\linewidth]{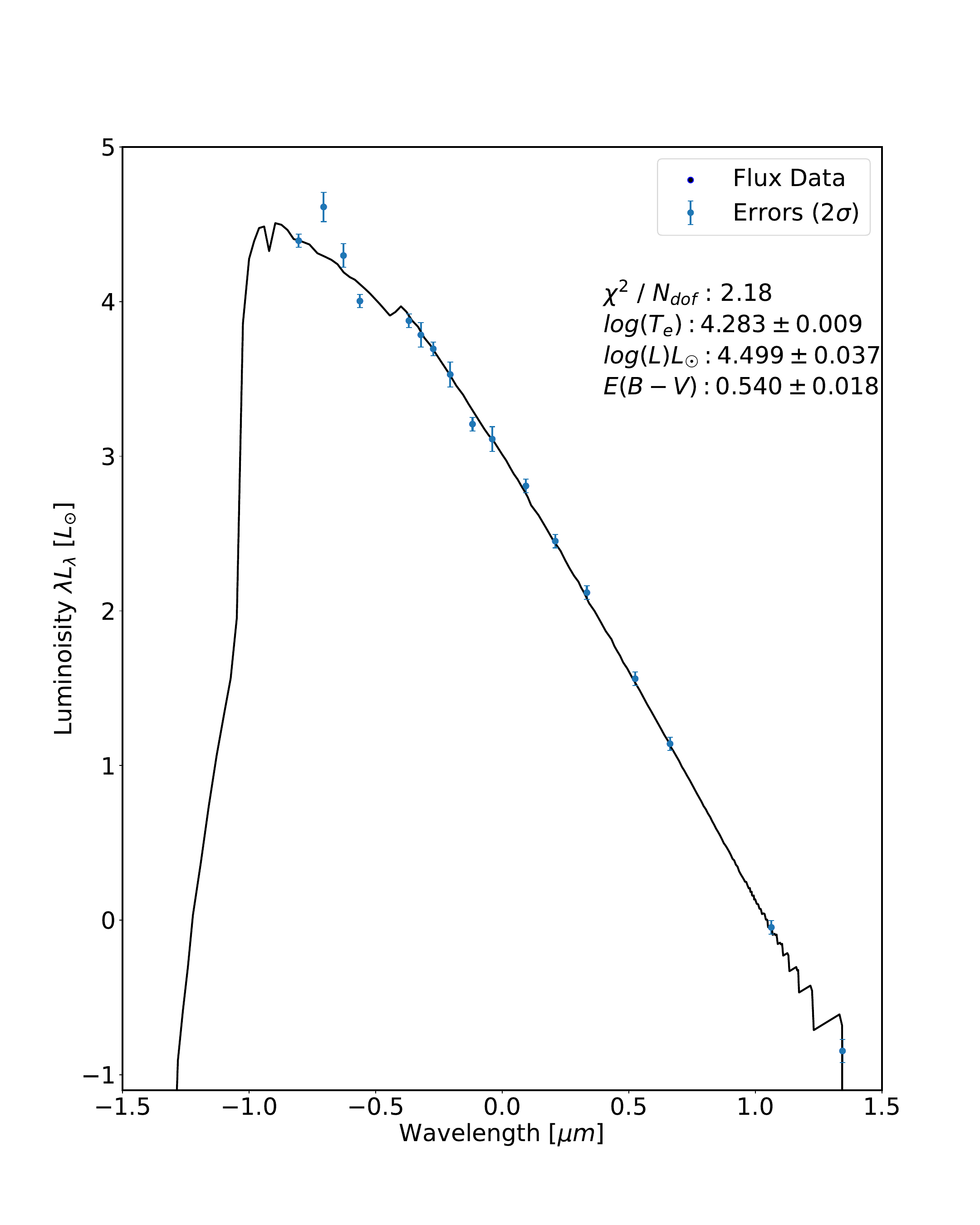}
    \caption{The SED of the B1III C star HD 36547.}
    \label{fig:star_3}
\end{figure}

\begin{figure}
    \includegraphics[width=\linewidth]{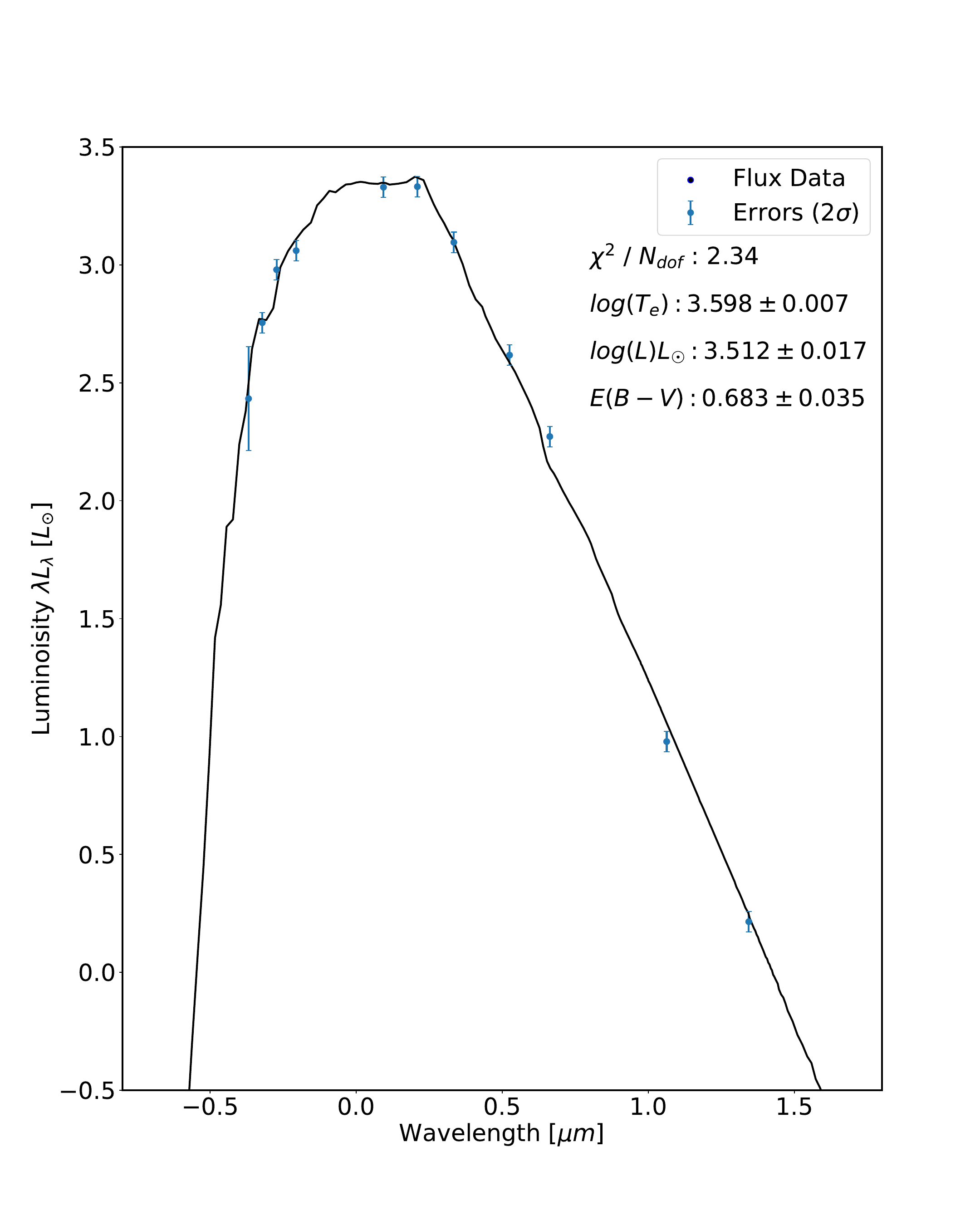}
    \caption{The SED of IRAS 5310+2411 which has no spectral classification.}
    \label{fig:star_4}
\end{figure}

\begin{figure}
    \includegraphics[width=\linewidth]{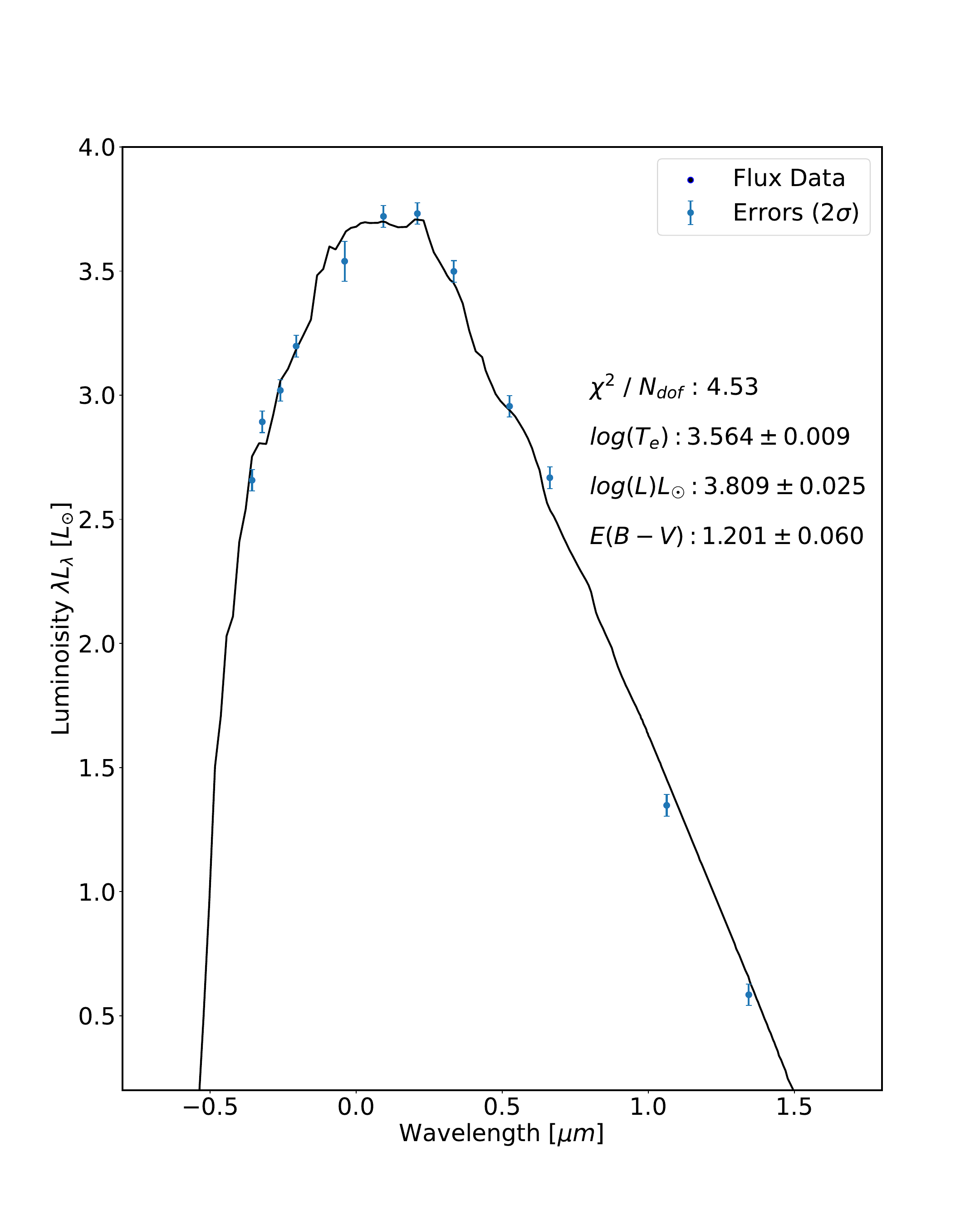}
    \caption{The SED of long period variable candidate star IRAS 5361-2406. It has no spectral classification.}
    \label{fig:star_5}
\end{figure}

Figures \ref{fig:star_1} through \ref{fig:star_5} show the SED fits for each of the 5 most luminous stars found near the Crab.


\section{Derivation of Equation 2} \label{appendix:B}
We start with Equation \ref{eq:1}, for a constant star formation rate of stars with masses $M>M_{min}$, and a mean mass of $\langle M \rangle = (x-1) M_{min}/(x-2)$. If we integrate over M, we get the number of stars

\[
    N =  \int_{t_{min}}^{t_{max}} \int_{M_{min}}^{\infty} \ dM \frac{dn}{dMdt} = \frac{SFR}{\langle M \rangle} \Delta t .
\]

\noindent For an interval of ages where $\Delta t = t_{max}-t_{min}$, the number of stars dying within a time period $\delta t$ is 

\[
    \int_{t_{min}}^{t_{max}} \ dt \int_{M(t)}^{M(t+\delta t)} \ dM \frac{dn}{dMdt} ,
\]
\noindent where $M(t)$ is the mass of a star dying at time t. If we do the mass integral, we get

\[
    \int_{t_{min}}^{t_{max}} \ dt \frac{SFR}{\langle M \rangle} \Bigg[ \Bigg(\frac{M(t)}{M_{min}}\Bigg)^{1-x} - \Bigg(\frac{M(t+\delta t)}{M_{min}}\Bigg)^{1-x}  \Bigg] .
\]

\noindent and then Taylor expand $M(t+\delta t) = M(t) + (dM/dt)\delta t$ to get 
\[
    \int_{t_{min}}^{t_{max}} \ dt \frac{SFR}{\langle M \rangle} \frac{dM}{dt} \Bigg(\frac{M(t)}{M_{min}} \Bigg)^{-x} (x-1)\delta t .
\]

\noindent We transform the integral, by change of variables, from time $t$ to mass $M$,

\[
    \int_{M(t_{min})}^{M(t_{max})} \ dM \frac{SFR}{\langle M \rangle} \Bigg( \frac{M(t)}{M_{min}} \Bigg)^{-x} (x-1) \delta t ,
\]
\noindent and solve the integral, which results in Equation \ref{eq:2}.


\end{appendix}



\bsp	
\label{lastpage}
\end{document}